\documentclass[a4paper, 12pt]{article}
\usepackage[cp1251]{inputenc}
\usepackage{amssymb}
\usepackage{amsmath}
\usepackage{amsfonts}
\usepackage{cite}
\usepackage{geometry}
\usepackage{graphicx}
\usepackage{afterpage}

\makeatletter
\renewcommand{\@biblabel}[1]{#1.\hfill}

\begin{document}

\begin{center}  	

	\textbf{\Large Acidity of graphene oxide aqueous solutions: \\ the role of hydroxyls unveiled} \\
	\vspace{5mm}
	\textbf{\large Yulia V. Novakovskaya} \\
	\textit{Lomonosov Moscow State University, 1/3, Leninskie Gory, Moscow, 119234 Russia \\
		E-mail: jvn@phys.chem.msu.ru}
	
\end{center}

\vspace{5mm}
	
	\textbf{Abstract}\\
Nonempirical modeling of hydrated single-layer graphene oxide (GO) fragments revealed that at least one, but substantial origin of the acidity of aqueous GO solutions is the dissociation in closely grouped and favorably H-bonded hydroxyls. The process can further be promoted by the adjacent $\pi$-conjugated carbon segments at the edges of GO flakes or their inner defects. The apparent $pK_a$ values of the groups fall in a range of ca. 2.0 to 5.3. Under certain conditions the ionization  may become spontaneous.

\vspace{10mm}

\section{Introduction}
\label{Introduction}

Nowadays, graphene oxide (GO) may seem too old to be of a valuable theoretical interest, the more so as the characteristics of the material vary depending on the method of synthesis \cite{GA}, which makes a task of elaborating a reasonable universal model of its structure practically unsolvable. Hence, any question related to the structure peculiarities seems ill defined. At the same time, there is a unique property of any GO specimen unclarified up to the present moment, namely, the high acidity of its aqueous solutions \cite{HH, CBH}. According to the titration data for different GO specimens, there should be about 1 acidic site per each 17 carbon atoms \cite{DAT}. This means that each tetrad of carbon hexagons should contain an acidic site with a roughly estimated $pK_a$ value of 3.93--3.96 \cite{DAT}, which is too high for typical carboxylic groups that are not as numerous in non-degraded GO specimens, and the more so for the prevailing individual hydroxyl groups \cite{GA}. As a kind of a compromise, vinylogous acids were suggested \cite{DAT} to be key groups in this respect, but to our mind the corresponding elegant dynamic model, which quite reasonably acquires for the disintegration of GO specimens in water, cannot account for the existence of numerous quite closely arranged acidic sites on non-degraded GO flakes almost immediately upon their formation, though the higher acidity of vinylogous acidic structural groups compared to usual carboxylic ones seems doubtless. 

To solve the problem one needs detailed information about the correlation between variations in the contents and local arrangements of oxygen-containing functional groups and the observed acidity of GO specimens. This by no means can be done in terms of purely experimental approach for nobody can state that not only the arrangements of functional groups but even the relative amounts of diverse groups are known. The qualitative and partly quantitative information of the latter kind can be retrieved from NMR, FTIR, or XP spectra, but even these data are far from unambiguity due to the variations in the peak positions of particular functional groups depending on their neighborhoods, including adjacent segments of carbon layer and water molecules \cite{FM,ZS,GN}.  Therefore, simulations of model systems, which involve diverse combinations of functional groups bound to the surface and edges of a carbon fragment, seem to be the only way that may help in clarifying the origins of the acidity of aqueous GO solutions. Constructing and analyzing diverse models enabled us to distinguish those where the acidity can be determined by closely arranged hydroxyl groups solely. And the result, though unexpected at first glance, but quite reasonable upon a closer view, is very interesting for the chemistry and chemical physics in general. 

To construct models, we took into account both the experimental data and theoretical concepts of GO structure that were in an excellent way summed up by Lerf, Dimiev, and Eigler \cite{GA}. In GO specimens prepared according to the most widely used original or modified Hummers techniques, the oxidized-to-graphitic carbon ratio varies from 61:39 \cite{Hs} to 69:31 \cite{T}. Areal estimates of the highly disordered oxidized regions, non-oxidized graphitic domains (up to 8 nm$^2$ in area), and defects, or holes (less than 5 nm$^2$) provided reasonably close values of 82, 16, and 2\%, respectively \cite{KVYK}. The surfaces of GO flakes are negatively charged as follows from the high absolute values of the negative $\zeta$-potential, which is nearly by half larger for the most deeply oxidized specimens \cite{KVYK}, and the progressively decreasing peak values in the cyclic voltammetry curves of the deeper oxidized specimens \cite{KVYK}. These observations additionally substantiate the relatively high dissociation constant of some functional groups in GO and prompts one to combine the basic ideas of Lerf and Klinowski \cite{LK}, Scholtz and Boehm \cite{SB}, Szabo and Dekany \cite{SD}, and Dimiev, Alemany, and Tour \cite{DAT}. 

\section{Simulation Methodology}
\label{Simulation Methodology}

To minimize the systems, basal carbon plane fragments were decorated chiefly with the groups that can act as proton donors or assist the process, which are carboxyls and carbonyls along edges and hydroxyls anywhere. Sin location of hydroxyls was not considered because the high inner strains that accompany their apperarance should inevitably lead to either rearrangement or bond-breakage reactions; and only anti hydroxyl pairs can be thought of as relatively long-lived.  The C/O ratio was varied approximately from 3:1 to 3:2, the latter one corresponding to a slightly larger oxygen content compared to the aforementioned averaged figures, but taking into account that we do not treat the graphitic and defect domains, and model strictly one layer (whereas experimental estimates reflect some averaged data for one and several-layer flakes), such variations in the integral composition are quite founded. Nonempirical simulations were carried out in the cluster approach because in real specimens there is no regularity and periodicity in the spatial arrangment of both oxidized and unoxidized carbon domains, as well as the decorating functional groups within the oxidized domains; and the effects related to the proton-donating ability seem to be of a relatively short-range nature. For solving the electronic problem, the density functional method with the B3LYP hydbrid functional was selected along with the double-$\zeta$ Gaussian basis set of 6-31G quality for basal plane carbon atoms and 6-31G(d,p) for all the residual atoms involved in functional groups and water molecules. Such combined basis set complies with the requirements to the sets used for simulating 2D or 3D objects (particularly, it is sufficiently compact to avoid linear dependence of basis functions when describing systems that involve up to two hundreds closely arranged atoms) and provides a balanced description of the grafted groups (it is sufficiently flexible due to the same large number of atoms, the tails of the atomic functions of which contribute to the description of the electron density distribution in their both more and less distant neighborhoods). The basal plane fragment involved 30 to 54 carbon atoms; the numbers of hydroxyl, carbonyl, and carboxyl groups were varied in ranges of 5--20, 1--3, and 1--5 respectively, and the number of water molecules was varied from 5 to 35. The correspondence of the structures to local minima of the adiabatic potential was confirmed by the normal-coordinate analysis; and thermal effects were tentatively taken into account by estimating the Gibbs energy increments (at 298K) determined solely by the vibrational degrees of freedom ($G_{vib}$), which is possible because differences between the $G_{rel}=E_e + G_{vib}$ relative values counted from the selected potential energy minimum (which are of primary importance to us) reflect only the changes related to some local internal structure transformations, when contributions of most of the degrees of freedom are nearly canceled out. Formal logarithmic equilibrium constants of the possible dissociation reactions ($pK_a$) were estimated based on $-\Delta G_{rel}/RT$ values. Simulations were carried out with the use of Firefly 8.2 package \cite{Firefly} and visualized with Chemcraft software \cite{Chemcraft}.

\section{Results and Discussion}
\label{Results and Discussion}

\begin{figure}[t] \center
	\includegraphics[width=1.0\textwidth]{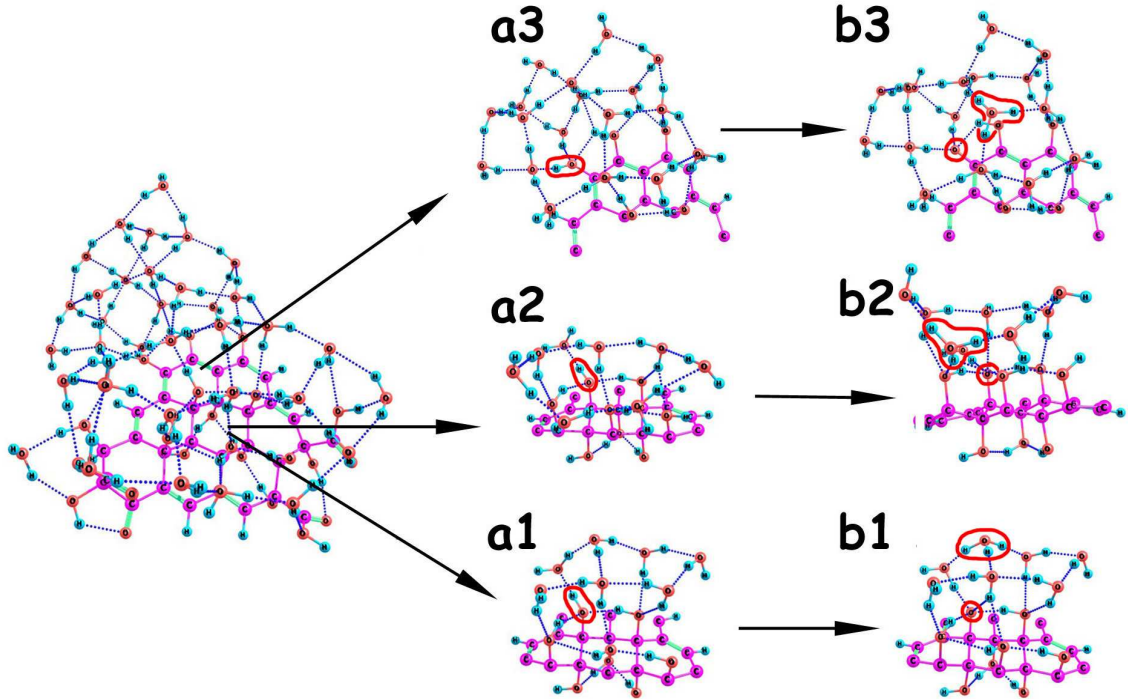}
	\caption{Hydrated graphene oxide model C$_{27}$H$_6$(O)(COOH)$_{3}$(OH)$_{15}$ $\cdot$ (H$_2$O)$_{35}$: (a1, a2, b1, b2) central and (a3, b3) edge segments of the model system (a1, a2, a3) before and (b1, b2, b3) after the dissociation of hydroxyls. Encircled are hydroxyl groups that can undergo dissociation and the resulting O$^{\delta -}$ and H$_3$O$^{\delta +}$ fragments.} \label{fig1} \vspace{5mm}
\end{figure}

Based on the results obtained, almost all the combinations of groups either arranged sequentially along an edge of a carbon fragment or over its plane either at a sufficient distance from or as close as possible to each other can be comparatively analyzed based on a C$_{30}$ carbon fragment. Furthermore, such a compact system can be supplemented with a sufficient number of water molecules to provide the formation of hydration spheres around the key structure segments. The spectacular compositions considered below can be represented by the following general formula: C$_{27}$H$_6$(O)(COOH)$_{3}$(OH)$_{15}$ $\cdot$ (H$_2$O)$_n$ with a varying number of water molecules, $n$=5--35. In the model GO fragment (which is sufficiently deeply oxidized to acquire a typical folded profile), the most of hydroxyls arranged over the basal plane and along its edges are grouped in diads, triads, and pentads. At the largest considered number of water molecules, $n$=35, there are a complete hydration monolayer over the functional groups at one side of the carbon plane and an extended hydration shell around one edge OH triad (Fig.\ref{fig1}). Hydroxyls, which reside in the central part of the carbon fragment, can form hydrogen bonds with each other so that water molecules are not included in the  corresponding H-bond sequence but rather attached to it from above. This H-bond subnetwork can either be continuous (a1 in Fig.\ref{fig1}) or broken (a2 in Fig.\ref{fig1}) depending on the preset orientation of the groups with respect to each other. According to the number of hydroxyls involved in a continuous H-bond subnetwork, their corresponding groups are referred to as either pentad or triad respectively. In both variants, there is one hydroxyl (encircled) that acts as an acceptor of two H-bond protons of the neighboring hydroxyls and a donor of its own proton in an H-bond to a water molecule. In the case of the edge hydroxyls (a3 in Fig.\ref{fig1}), their larger freedom in the relative orientation results in the formation of an H-bond network, in which they are interspersed with water molecules though a certain orientation again toward one OH group (encircled) can be noticed. It is these encircled hydroxyls that are primary candidates for the dissociation; and the simulations proved this idea.

In the case of the hydroxyl in the central part of the carbon fragment, a stable configuration that involves a  water separated (C)O$^-$\dots (H$_2$O)\dots H$_3$O$^+$ ion pair was found (b1 in Fig.\ref{fig1}); and its adiabatic potential energy is only 2.5 kcal/mol higher than that of the original neutral structure. When thermal energy increments are taken into account, the resulting difference between the Gibbs energies becomes only slightly larger, 2.7 kcal/mol, which formally corresponds to the $pK_a$ value of the hydroxyl dissociation of 2.0. This value is very high due to the stabilization of the (C)O$^-$ group provided by the ordered H-bond sequences of hydroxyls that converge to this very oxygen in a nearly coplanar ring-like structure, which is stable due to the coordinating carbon footing. It is worth noting that the above $pK_a$ value corresponds to the initial stage of the dissociation. At a larger amount of water molecules, the energy of the system can become even lower due to the additional stabilization predetermined by the larger spatial separation of charges.

When the close neighborhood of the same hydroxyl is seemingly only slightly changed, namely, the spatial orientation of two hydroxyls separated from this one by at least one additional group is changed so that they are no longer involved in the H-bond sequences that end up at the key hydroxyl group, and the immediate neighborhood of the latter comprises only two rather than four hydroxyls oriented toward it, the situation changes drastically. This central hydroxyl of the triad is not inclined to lose a proton to form a water separated (C)O$^-$\dots (H$_2$O)\dots H$_3$O$^+$ ion pair with the involvement of two neighboring molecules. Stabilization of the charge-separated configuration was found to be provided only when the (C)O$^-$ and H$_3$O$^+$ fragments were separated with a hydroxyl group, which prevented the formation of a direct short path for the sequential proton shifts that would internally neutralize the configuration (b2 in Fig.\ref{fig1}). And such a structure has an energy higher by 9.2 kcal/mol than the corresponding all neutral one. It should be noted that the reorientation itself of two hydroxyls, which made the H-bonded segments within the discussed part of the system shorter at the same total number of H-bonds, resulted in an increase in the total energy of the system by 2.8 kcal/mol. As to the formation of the zwitterionic configuration, the thermal increments lowered its apparent relative free energy to 6.8 kcal/mol, which corresponds to the formal equilibrium constant of the ionization reaction $pK_a$ = 3.5. 

In the case of the edge hydroxyl, the situation is visually similar but quantitatively different. The group can also lose a proton to produce a more distant hydronium ion, the shortest separations of which involve three water molecules: (C)O$^-$\dots (H$_2$O)$_3$\dots H$_3$O$^+$ (b3 in Fig.\ref{fig1}). Again, the (C)O$^-$ group acts as a tri-coordinate knot in the H-bond network, but here the stabilization of O$^-$ is provided by its involvement in the 10-atomic $\pi$-conjugated carbon segment, which should produce a stronger effect compared to the above H-bond network solely. And the strength is unexpectedly high: the adiabatic potential energy of the dissociation is negative! It equals $\sim$-4.9 kcal/mol, and $\Delta G_{rel}$, though smaller in absolute value, is also negative (-2.5 kcal/mol), which means that the dissociation of such edge hydroxyl groups is thermodynamically favorable, and the corresponding formal equilibrium constant is of an order of $10^2$. 

\begin{figure}[t] \center
	\includegraphics[width=1.0\textwidth]{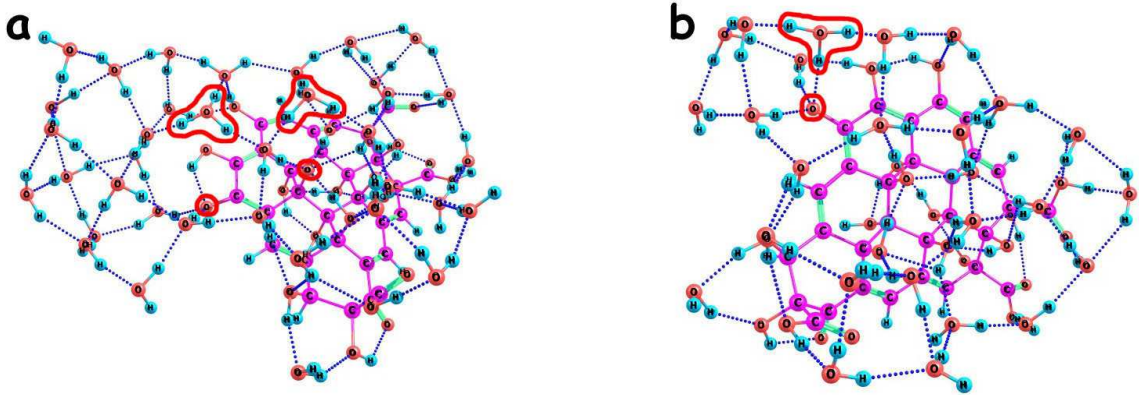}
	\caption{Hydrated graphene oxide model C$_{27}$H$_6$(O)(COOH)$_{3}$(OH)$_{15}$ $\cdot$ (H$_2$O)$_n$ at (a) $n=35$, side view and (b) $n=25$, top view. Encircled are the existing O$^{\delta -}$ and H$_3$O$^{\delta +}$ fragments.} \label{fig2} \vspace{5mm}
\end{figure}

Furthermore, the complementary stabilization of the two dissociated hydroxyl groups, one at the edge and another, over the basal carbon plane, makes their coexistence energetically possible. Even within the small model fragment we consider, a configuration where both groups lost their protons to form water separated ion pairs was found (Fig.\ref{fig2}a), which complies with the aforementioned estimate of one acidic site per each tetrad of carbon rings. Remarkably, the adiabatic potential energy of this double zwitterionic configuration is only 4.5 kcal/mol higher than that of the all-neutral structure, and $\Delta G_{rel}$=4.8 kcal/mol.  

The a priori unexpected thermodynamic favorableness of the dissociation of an edge hydroxyl has brought us to an idea of analyzing the process in more detail. When the number of water molecules in the model system was $n$=25, a spontaneous dissociation of the group was found to take place, and a (C)O$^-$\dots H$_3$O$^+$ contact ion pair appeared (Fig. \ref{fig2}b). A reason for that may be the following. At this number of water molecules, the edge hydroxyl triad is yet similar to the one over the carbon plane in that these hydroxyls form a directional H-bonded sequence, but it is not parallel to the carbon footing, rather coplanar to the aforementioned conjugated 10-atomic segment. As a result, here the H-bond correlation is supplemented with a $\pi$-conjugation within the subsystem of second-row elements (basal carbons and an oxygen anion), which stabilizes the (C)O$^-$ group, and, hence, assists in the dissociation. 

\section{Conclusions}
\label{Conclusions}

Thus, within at least triads of hydroxyls, where two groups as proton donors assist the third one, this latter can dissociate irrespectively of whether it is located at the edge or over the basal carbon plane of a GO flake. The process may become spontaneous at a moderate hydration when the hydroxyls are bound to an incompletely oxidized edge of a carbon sheet. The dissociation of a hydroxyl located on a basal carbon plane can be characterized by apparent $pK_a$ values in a range of 2.0 to 3.5 in the absence of the neighboring dissociated groups and up to 5.3 when there is already a (C)O$^-$\dots (H$_2$O)$_k$\dots H$_3$O$^+$ ion pair in its relatively close proximity. Seemingly minor changes in the mutual orientation of hydroxyls formed over the carbon sheet as a result of oxidation can substantially change the apparent dissociation constant. Remarkably, when the amount of water is moderate, and either two H-bonded sequences of hydroxyls end up at the same hydroxyl group, which, hence, acts as a double proton acceptor in its H-bonds with the neighboring hydroxyls and a proton donor in an H-bond to a water molecule, or one similar H-bonded sequence of hydroxyls is formed nearby a $\pi$-conjugated carbon segment, the hydroxyl can dissociate. The longer the H-bonded sequences, the higher the dissociation constant. Furthermore, judging from the simulation results obtained, the stronger the correlation between the functional groups and the carbon footing, the more probable the dissociation process, which means that it should proceed more easily already at the stage of separation of GO flakes. Thus, by contrast to usual trends in dissociation, which is facilitated by the increasing amount of water, here the ordering and correlation of OH groups in H-bond network and the existence of adjacent $\pi$-conjugated carbon segments play the more substantial role. Note that the two hydration spheres around the edge hydroxyl and the average distance between the mean carbon and water planes of ca. 5.1 \AA\ in our simulations correspond to the GO layer separation of no smaller than 9 \AA. A subsequent increase in the amount of water seems to provide only a larger separation of charges due to the migration of hydronium ions. 

We would like to stress that the goal of the study was not to consider all the possible groups that can be ionized during the formation and operation of graphene oxide species in aqueous solutions but rather to clarify whether hydroxyls that are most abundant modifying groups of GO flakes can substantially contribute to the observed acidity of the material. And the positive answer to this question seems to be of a principal importance to the chemical physics of not only GO, but any extended carbon-based highly hydroxylated systems.

\end{document}